\documentclass[proceedings,preprint]{rmaa}

\usepackage{paralist}
\usepackage{psfrag,color}

\SetYear{2011}
\SetConfTitle{LARIM}

\title{Solar origins: Place and Chemical Composition} 

\author{
Leticia Carigi\altaffilmark{1} and  Manuel Peimbert\altaffilmark{1}}

\altaffiltext{1}{Instituto de Astronom\'ia, Universidad Nacional Aut\'onoma de M\'exico, 
A.P. 70264, D.F., 04510 M\'exico. (carigi, peimbert@astroscu.unam.mx).}
\shortauthor{Carigi \&  Peimbert}

\shorttitle{Sun migration and chemical evolution}

\listofauthors{L. Carigi \& M. Peimbert}
\indexauthor{Carigi, L.}
\indexauthor{Peimbert, M.}

\abstract
{
We discuss a chemical evolution model with 
$Z$-dependent yields that reproduces the O/H, C/H, and C/O 
gradients of the Galactic disk and the chemical history of the solar vicinity.
The model fits the H, He, C, and O abundances derived from
recombination lines of the H~{\sc ii} region M17 (including the fraction of C    
and O atoms embedded in dust); the protosolar H,
He, C, O, and Fe abundances; and the C/O-O/H, C/Fe-Fe/H, and O/Fe-Fe/H
relations derived from stars of the solar vicinity.  The agreement
of the model with the protosolar abundances at the Sun-formation time
implies that the Sun originated from a well mixed ISM
at a galactocentric distance of $7.6 \pm 0.8$ kpc. 
}

\resumen
{
Discutimos un modelo de evoluci\'on qu\'imica con rendimientos dependientes de
$Z$ que reproduce los gradientes de O/H, C/H y C/O del disco Gal\'actico y la
historia qu\'imica de la vecindad solar.
El modelo ajusta las abundancias de H, He, C y O derivadas a partir de l\'ineas
de recombinaci\'on en la regi\'on H~{\sc ii} M17 (incluyendo
la fracci\'on de \'atomos de C y O embebida en polvo); 
las abundancias protosolares de H, He, C, O y Fe;
y las relaciones C/O-O/H, C/Fe-Fe/H y O/Fe-Fe/H derivadas de estrellas de la vecindad solar.
El ajuste del modelo con las abundancias protosolares al tiempo de formaci\'on 
del Sol implica que el Sol se origin\'o en un medio qu\'imicamente bien mezclado
a una distancia galactoc\'entrica de $7.6 \pm 0.8$ kpc.
}

\addkeyword{galaxies: chemical evolution}
\addkeyword{H~II regions: M17, Orion nebula} 
\addkeyword{ISM: abundances} 
\addkeyword{Sun: abundances}

\begin{document}
\maketitle


\section{Introduction}
\label{sec:intro}

The comparison of detailed Galactic chemical evolution models, GCE models,
with accurate abundance determinations of stars and gaseous
nebulae provides a powerful tool to test the GCE
models and the accuracy of observational abundance determinations of
stars of different ages and of H{\sc~ii} regions located at different
galactocentric distances.

To find a robust model for the Galaxy we decided to use as 
main observational constraint the slope and absolute value of the O/H
gradient. This model has been tested with other observational constraints
of high quality. 
A full discussion of this problem on the
GCE has been presented elsewhere (Carigi and Peimbert 2011). 

\section{Discussion}
\label{sec:discussion}

GCE models can be constrained by chemical gradients obtained by
different methods used in the literature.
Those Galactic disk gradients in general 
show similar slopes but a considerable spread in the absolute O/H ratios.
A comparison of  many of the different methods used has
been made by \citet{kew08}. They find that the O/H differences
derived by different methods for a given H{\sc~ii} region
can be as large as 0.7 dex.
Most of the differences among the various calibrations are due to the temperature
distribution inside the nebulae. 
The recombination lines (RL) method, that is almost
independent of the electron temperature,
produces gaseous O and C abundances higher
by about 0.15 to 0.35 dex than the forbidden lines (FL) method, that is strongly
dependent on the electron temperature.
It is possible to increase the
FL abundances under the assumption of temperature inhomogeneities 
to reach agreement with the RL values \citep{pea11}.

In this work we use  H, He, C, and O abundances of H{\sc~ii} regions 
at different galactocentric distances based only on the RL method.
Since the recombinatio line ratios among these
four elements are practically independent of the electron temperature their
relative abundances are very reliable.

Moreover, these abundances were corrected under the assumption that 35\% of 
the O atoms and 25\% of the C atoms are trapped in dust grains \citep{pea10},
increasing the O/H and C/H values by 0.12 and 0.10 dex, respectively.
In order to constrain the chemical history of the interstellar medium, 
we considered stellar abundances of different ages.

In Figure 1 we present our model together with the best observational constraints
available (those listed above).
The model was built to reproduce the O/H gradient (upper right panel) 
assuming an inside-out scenario and intermediate wind yields 
(moderate mass loss rate for massive stars of $Z_\odot$).
Specifically our model is able to reproduce the following observational 
constraints presented in the other four panels: a) the current O/H, C/H, and C/O
abundance gradients (slopes and absolute values) derived from Galactic
H~{\sc{ii}} regions;
b) the O/H, C/H, and Fe/H values and Fe/H-time relation derived 
from halo and disk stars of different ages in the solar vicinity;
c) the He/H, C/H, O/H, and Fe/H protosolar abundances
that correspond to those present in the interstellar medium 4.5 Gyr ago;
and
d) the He/H, C/H, and O/H values of the Galactic H~{\sc{ii}} regions M17 and M20.

From the protosolar 12 + log(O/H) = 8.73 $\pm 0.05$ value by Asplund et al. (2009)
and the O/H values predicted by our GCE model for $r = 8$ and $r = 7$ kpc
of 8.70 and 8.77 for the ISM when the Sun was formed ($t=8.5$ Gyr),
we obtain that the Sun originated at $r = 7.6 \pm 0.8$ kpc.

We also decided to compare our best models with the C/O vs O/H results
derived by  \citet{est02,est09} from bright H{\sc~ii} regions in nearby
spiral galaxies based on recombination lines and including the dust correction.

\begin{figure}[!t] \includegraphics[width=\columnwidth]{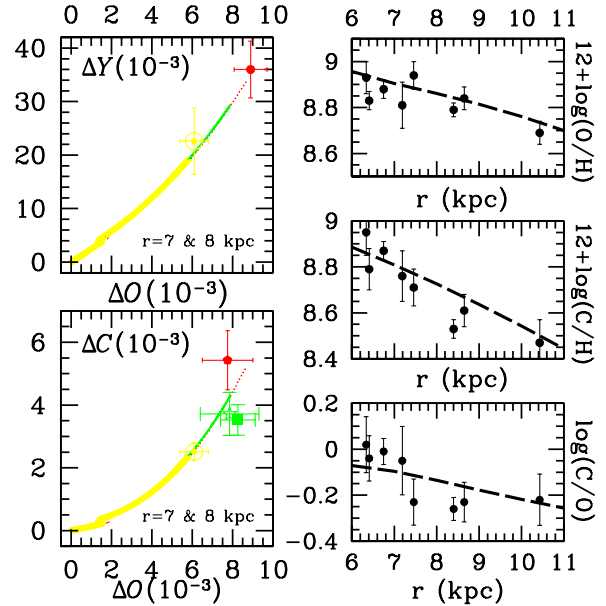}
\caption{ Chemical evolution model for the solar vicinity and the Galactic disk. 
The left panels show the 0-13Gyr evolution of $\Delta Y$ vs $\Delta O$
and of $\Delta C$ vs $\Delta O$ for $r = 7$ kpc (dotted red lines)
and $r = 8$ kpc (thin solid green line), the
thick yellow lines show the evolution from 0 to 8.5 Gyr for $r$ = 8 kpc. 
The right panels show the present-day ISM abundance
ratios as a function of galactocentric distance.
Data: 
{\it Filled red pentagon}: Average
values of the M17 and M20 H~{\sc ii} regions from \citet{car08} at $r=$ 6.75 and 7.19 kpc,
respectively.
{\it Yellow $\odot$}: Protosolar values by \citet{asp09}.
{\it  Empty green star}: Average values of young F-G dwarf stars
of the solar vicinity from \citet{ben06}.
{\it Filled green square}: Average values of NGC 3576 and Orion H~{\sc ii}
regions at  $r=$ 7.46 and 8.40 kpc.
{\it Filled black circles}: H~{\sc ii} regions \citep{gar07},
gas \citep{gar07} plus dust correction \citep{pea10}.
}
\label{fig:GradientIntermediate} \end{figure}

\section{Conclusions} \label{sec:conclusions}

The agreement of the He/O, C/O, and Fe/O ratios between the model and the
protosolar abundances implies that the Sun formed from a well mixed ISM,
since about half of the freshly He (and C) is produced
by massive stars and half by low-and-intermediate-mass stars \citep{car05,car08}, 
and an important fraction of the Fe comes from SNIa. 

The agreement of our model with the protosolar abundances
and the Sun-formation time supports the idea that the Sun originated
at $7.6 \pm 0.8$ kpc, close to its current galactocentric distance ($r=8$ kpc).

We obtain that our chemical evolution model of the Galactic disk
for the present time
produces a reasonable fit to the O/H-C/O relation
derived from H~{\sc{ii}} regions of nearby spiral galaxies 
This agreement might imply  that spiral galaxies have a similar IMF, 
no selective outflows,
and probably a formation scenario similar to that of our galaxy
(for a more general discussion of this issue see Carigi \& Peimbert 2011).

\end{document}